# Transverse ultrasound absorption in cubic crystals with positive and negative anisotropies of second-order elasticity moduli

I.G. Kuleyev, I.I. Kuleyev and I.Yu. Arapova

Institute of Metal Physics, Ural Division, Russian Academy of Sciences, 620219 Ekaterinburg, Russia

e-mail: kuleev@imp.uran.ru

**Abstract**

The transverse ultrasound absorption in cubic crystals with positive and negative anisotropies of the second-order elasticity moduli is analyzed. The scattering of the ultrasound by defects and during anharmonic scattering processes is considered. The transverse ultrasound absorption is analyzed as a function of the wave-vector direction in terms of the anisotropic continuum model. The Landau-Rumer mechanism is considered for anharmonic scattering processes. Known values of the second- and third-order elasticity moduli are used to calculate parameters determining the ultrasound absorption. It is shown that the angular dependences of the transverse ultrasound absorption differ qualitatively if the anharmonic scattering processes dominate in cubic crystals with positive and negative anisotropies of the second-order elasticity moduli. For the scattering by defects and the anharmonic scattering processes, the angular dependences of the transverse ultrasound absorption exhibit the inverse behavior, making it possible to determine the dominating mechanism of the ultrasound relaxation in the crystals under study.

## 1. Introduction

A problem encountered in studies of the ultrasound absorption [1-4], the phonon transport [5-6] and the thermal e.m.f. of the electron-phonon drag [7] in semiconductor and dielectric crystals is the correct consideration of the effect of the cubic anisotropy on the spectrum, the oscillation mode polarization and the phonon relaxation rates. When the temperatures are sufficiently low and the inequality $\omega_{q\lambda} \tau_{ph}^{\lambda}(\mathbf{q},T) >> 1$ is fulfilled ($\tau_{ph}^{\lambda}(\mathbf{q},T) = 1/\nu_{ph}^{\lambda}(\mathbf{q},T)$, $\nu_{ph}^{\lambda}(\mathbf{q},T)$ being the phonon relaxation rate and $\omega_{q\lambda}$ the frequency of phonons with the wave-vector *q* and the polarization *λ*), the dominant contribution to the volume absorption of sound waves is due to the scattering by defects, including isotopic scattering, and normal processes of the phonon-phonon scattering (see, e.g., [1, 4]). We shall restrict ourselves to their consideration in this paper. The isotropic medium approximation [4-6, 8-11], which is commonly used to estimate the probability of various scattering processes, is inadequate for germanium, silicon, diamond and other semiconductor crystals having the cubic symmetry and a considerable anisotropy both of the second- and third-order elasticity moduli. The anisotropic continuum model provides a convenient solution to these problems. In this model, the harmonic energy of cubic crystals is expressed as three second-order elasticity moduli, while the anharmonic energy as six third-order elasticity moduli. Notice that the second- and third-order elasticity moduli have been determined experimentally for a considerable number of cubic crystals. Therefore, the phonon relaxation rates calculated in terms of this model present a reliable basis for the interpretation of experimental data on the ultrasound absorption and the phonon transport in cubic crystals.

Using experimentally determined elasticity moduli of the second and third orders, the researchers [12] derived an equation for the elastic energy associated with the lattice anharmonicity in cubic crystals and calculated phonon relaxation rates for some anharmonic scattering processes. The transverse ultrasound absorption in cubic Ge, Si and diamond crystals was discussed in [13]. If the cubic anisotropy in the matrix element describing three-phonon scattering processes was taken into account, the dependences of the relaxation rates on the wave-vector of transverse phonons were shown to be qualitatively different from the classical linear Landau-Rumer dependence [14]. However, the analysis [12, 13] was performed in the isotropic approximation for the spectrum and the polarization of phonons: the effect of the cubic anisotropy on the phonon spectrum was disregarded in the energy conservation law. Moreover, the vibrational modes were assumed to be pure (purely longitudinal or purely transverse) modes. As is known [1, 15, 16], quasi-longitudinal or quasi-transverse vibrations propagate in cubic crystals, while pure modes propagate only in symmetric directions such as [100], [110] and [111]. The analysis of the spectrum and the polarization vectors of vibrational branches [17] demonstrated that all cubic crystals can be classed into those with the positive ($\Delta C>0$) and the negative ($\Delta C<0$) anisotropy of the second-order elasticity moduli depending on the sign of $\Delta C$ ($\Delta C=c_{12}+2c_{44}-c_{11}$, where $c_{ij}$ denotes the second-order elasticity moduli) (see [17], Table 1). This parameter is zero in isotropic media. The first type ($\Delta C>0$) includes Ge, Si, diamond, IbSb and GaSb crystals. KCl, NaCl and GaAs crystals are referred to the second type ($\Delta C<0$). The form of the spectra of vibrational branches and the angular dependences of polarization vectors of quasi-transverse modes differ qualitatively in crystals of the first and second types; these characteristics differ just quantitatively in crystals of one type [17]. The question arises of how this difference influences the behavior of the relaxation characteristics of cubic crystals: Will the angular dependences of the ultrasound absorption in the crystals of the first and second types have qualitative differences? In other words, whether the ultrasound absorption anisotropy will be determined predominantly by the anisotropy of the harmonic or the anharmonic energy.

Such analysis was performed [18] for the relaxation rates of transverse phonons in terms of the Landau-Rumer mechanism. Unlike earlier calculations, the analysis took into account exactly the effect of the cubic anisotropy on both the spectrum of the vibrational branches involved in the energy conservation law and the phonon polarization vectors. The contributions of the longitudinal components of quasi-transverse vibrations to the phonon relaxation rates were estimated for the two types of cubic crystals. It was shown that the quasi-isotropic approximation is not adequate for the quantitative description of the anisotropy of the relaxation rates of quasi-transverse phonons.

In what follows we shall analyze the angular dependences of the quasi-transverse ultrasound absorption in cubic crystals of the two types for the scattering by defects and in terms of the Landau-Rumer mechanism. As to the phonon polarization vectors, we shall restrict ourselves to pure modes for the quasi-longitudinal vibrations. It was shown [17] that the error induced by this approximation is small for longitudinal phonons: less than 1% in Ge, Si and diamond crystals, and 3% in KCl. In the case of quasi-transverse phonons moving in directions other than symmetric ones, the maximum contribution of the longitudinal component to the transverse-longitudinal vibrations is 27% in KCl, 16.5% in Ge, 10% in Si, and 8% in diamond [17]. Therefore, our calculations of the ultrasound absorption in an arbitrary direction of the wave vector will allow for the contribution of the longitudinal component to the transverse-longitudinal vibrations similarly to [18]. It will be shown that the angular dependences of the absorption of quasi-transverse vibrational modes are qualitatively different in cubic crystals with the positive and the negative anisotropy of the second-order elasticity moduli.

## 2. Absorption of Transverse Ultrasound in Cubic Crystals Subject to Competitive Scattering on Defects and Anharmonic Scattering Processes

If the inequality $\omega_{q\lambda}\tau_{ph}^{\lambda}(\mathbf{q},T) >> 1$ ($\tau_{ph}^{\lambda}(\mathbf{q},T) = 1/\nu_{ph}^{\lambda}(\mathbf{q},T)$ and $\omega_{q\lambda}$ being the frequency of a phonon with a wave-vector $q$ and a polarization $\lambda$) is fulfilled, the ultrasonic wave absorption $\alpha_\lambda(\mathbf{q})$ with a wave-vector $q$ and a polarization $\lambda$ is proportional to the phonon relaxation rate $\nu_{ph}^\lambda(\mathbf{q},T)$ (see, e.g., [1-3]):

$$\alpha_\lambda(\mathbf{q},T) = \frac{\nu_{ph}^\lambda(\mathbf{q},T)}{2 S_\lambda(\mathbf{q})}, \tag{1}$$

where $S_\lambda(\mathbf{q}) = S_\lambda(\theta,\varphi)$ is the phonon phase velocity, which depends on the angular variables $\theta$ and $\varphi$ of the vector $\mathbf{q}$, and $T$ is the temperature. Experimental studies of the ultrasound absorption [1, 19] demonstrated that the inequality $\omega_{q\lambda}\tau_{ph}^\lambda(\mathbf{q},T) \gg 1$ is fulfilled at sufficiently low temperatures. For example, it holds at temperatures below 50 K, 100 K and 300 K for germanium, silicon and diamond crystals respectively. Therefore, in what follows we shall only consider the intervals of temperatures and wave vectors $\mathbf{q}$ when this inequality is fulfilled. In this case, the dominant contribution to the volume absorption of acoustic waves is due to the scattering by defects, including the isotopic scattering, and normal processes of the phonon-phonon scattering (see, e.g., [1, 10]). We shall restrict ourselves to the consideration of these relaxation processes while analyzing the absorption of the long-wave transverse ultrasound ($\hbar\omega_{qt} \ll k_B T$).

In accordance with [20, 21], the expression for the phonon relaxation rate subject to the elastic scattering by point defects in cubic crystals is

$$\nu_{phi}(q_1,\lambda_1) = \pi G V_0 \omega_{q_1\lambda_1}^2 \frac{1}{2V}\sum_{\mathbf{q}_2\lambda_2} \delta(\omega_{q_1\lambda_1} - \omega_{q_2\lambda_2})\left|\mathbf{e}_{\mathbf{q}_1\lambda_1}\mathbf{e}^*_{\mathbf{q}_2\lambda_2}\right|^2. \tag{2}$$

Here $V_0$ is the volume per one atom, $\mathbf{e}_{\mathbf{q}\lambda_1}$ is the phonon polarization vector, and $G$ characterizes the scattering intensity. In the case of the scattering by the isotopic disorder or impurities of the concentration $N_i$, the scattering intensity $G$ has the form [19-22]

$$G = g = \sum_i C_i \left(\frac{\Delta M_i}{\overline{M}}\right)^2, \quad u \quad G = V_0 N_i\left[\sigma_1^2 + (\sigma_2 + \sigma_3)^2\right], \tag{3}$$

where $g$ is the isotopic disorder factor, $\Delta M = M_i - \overline{M}$, $M_i$ is the mass of the $i$-th isotope, $\overline{M} = \sum_i C_i M_i$ is the average mass of the isotope composition, $C_i$ is the concentration of the $i$-th isotope, while $\sigma_1$, $\sigma_2$ and $\sigma_3$ characterize the contributions from the change of the unit cell mass, the force constants and the lattice deformation to the scattering cross-section of phonons by impurities (for more information see [5, 19, 21]). Expression (2) can be written in the form

$$\nu_{phi}(q_1,\lambda_1) = \frac{GV_0 \omega_{q_1\lambda_1}^4}{16\pi^2} \Phi_{q_1\lambda_1}(\theta_1,\varphi_1), \quad \Phi_{q_1\lambda_1}(\theta_1,\varphi_1) = \sum_{\lambda_2}\int_{-1}^{1}dx\int_{0}^{2\pi}d\varphi_2 \frac{\left|\left(\mathbf{e}_{\mathbf{q}_1}^{\lambda_1}\mathbf{e}_{\mathbf{q}_2}^{*\lambda_2}\right)\right|^2}{(S_{\lambda_2}(\theta_2,\varphi_2))^3}. \quad (4)$$

It is easy to show for the isotropic medium that

$$\Phi_{q_1\lambda_1}(\theta_1,\varphi_1) = \frac{4\pi}{3}\left(\frac{1}{S_L^3} + \frac{2}{S_t^3}\right), \quad (5)$$

where $S_L$ and $S_t$ are velocities of longitudinal and transverse phonons. Using the data on phase velocities and polarization vectors of phonons [17], we found that in cubic crystals the function $\Phi$ does not depend on either the polarization or the scattering angles of phonon:

$$\Phi_{q_1\lambda_1}(\theta_1,\varphi_1) = \Phi = \frac{4\pi}{3}\left(\left\langle\frac{1}{S_L^3}\right\rangle + \left\langle\frac{1}{S_{t1}^3}\right\rangle + \left\langle\frac{1}{S_{t2}^3}\right\rangle\right), \quad \left\langle\frac{1}{S_\lambda^3}\right\rangle = \frac{1}{4\pi}\int d\Omega \frac{1}{(S_\lambda(\theta,\varphi))^3}. \quad (5a)$$

The values of $\Phi$ for the crystals under study are given in Table 2. So, the relaxation rates of phonons during their elastic scattering by defects in cubic crystals are equal for longitudinal and transverse phonons and depend only on the scattered phonon energy:

$$\nu_{phi}(q_1,\lambda_1) = \frac{\pi}{6}GV_0\omega_{q_1\lambda_1}^2 D(\omega_{q_1\lambda_1}), \quad D(\omega_{q_1\lambda_1}) = \frac{1}{V}\sum_{\mathbf{q}_2\lambda_2}\delta(\omega_{q_1\lambda_1}-\omega_{q_2\lambda_2}) = \frac{3}{(2\pi)^3}(\omega_{q_1\lambda_1})^2\Phi. \quad (6)$$

Expression (6) can be written in the form

$$\nu_{phi} \cong B_i T_q^4, \quad T_q \equiv \hbar\omega_{qt}/k_B, \quad B_i = G(\hbar/k_B)^4(V_0\Phi/8\pi^2), \quad (6a)$$

where $T_q$ is the ultrasonic quantum energy in Kelvin. Expression (6) for the scattering **by the** isotopic disorder in cubic crystals coincides with the equation derived in [20, 21].

In accordance with the prevailing concepts [1-6, 8-11], the main mechanism of the transverse phonon relaxation in normal three-phonon scattering processes is the Landau-Rumer mechanism [14] when the merging of the transverse and the longitudinal phonon produces a longitudinal phonon (T+L→L). The expression for the relaxation rate of phonons moving in an arbitrary direction relative to the crystallographic axes of crystals was obtained for this mechanism in terms of the anisotropic continuum model in [18]. Unlike earlier calculations, this expression exactly allows for the effect of the cubic anisotropy both on the spectrum of the vibrational branches involved in the energy conservation law and on the phonon polarization vectors. In the long-wave approximation, at temperatures much lower than the Debye temperature the relaxation rate has the form [18]:

$$\nu^{TLL}(\theta_1,\varphi_1) = B_0^{TLL}(\theta_1,\varphi_1)T_q T^4, \quad B_0^{TLL}(\theta_1,\varphi_1) = B_0 J(\theta_1,\varphi_1), \quad B_0 = \frac{\pi^3 k_B^5}{15\hbar^4 \rho^3 S_t(\theta_1,\varphi_1)\langle S_L\rangle^8}, \quad (7)$$

where $J(\theta_1,\varphi_1) = \int_{-1}^{1} dx \frac{1}{\pi} \int_{0}^{2\pi} d\varphi_2 \delta(\cos\theta_{12} - S^{**}) \frac{I(\theta_2,\varphi_2,\theta_1,\varphi_1)}{(1+\Delta_2)^8}$, $x = \cos\theta_2$,

$$S^{**} = \lim_{y \to 0}\left[S^*(1-\Delta_2) + (\Delta_2 - \Delta_3)/y\right]. \quad S^* = S_t(\theta_1,\varphi_1)/S_L \qquad (8)$$

$\cos\theta_{12} = \sin\theta_1 \cos(\varphi_2 - \varphi_1)\sin\theta_2 + \cos\theta_1 \cos\theta_2$.

Here $\rho$ is the density, while the variables $\theta_1$ and $\varphi_1$ determine the direction of the sound wave vector $\mathbf{q_1}$ relative to the crystallographic axes. The spectrum and the polarization vectors of the vibration modes in an arbitrary direction of the wave vector were determined in a system of coordinates connected with the cube edges. The quantities $\Delta_2^L(\theta_2,\varphi_2)$ and $\Delta_3^L(\theta_3,\varphi_3)$ characterize the anisotropy of the phase velocity of longitudinal phonons with the wave vectors $\mathbf{q_2}$ and $\mathbf{q_3}$:

$$\Delta_2^L(\theta_2,\varphi_2) = (S_2^L(\theta_2,\varphi_2) - 1)/S_L, \quad S_L = \langle S_L(\theta,\varphi)\rangle = \frac{1}{4\pi}\int d\Omega S_L(\theta,\varphi). \qquad (9)$$

Considering the law of phonon momentum conservation, the angular variables $\theta_3$ and $\varphi_3$ of the vector $\mathbf{q_3}$ are expressed as the angular variables $\theta_1$, $\varphi_1$ and $\theta_2$, $\varphi_2$ of the wave vectors $\mathbf{q_1}$ and $\mathbf{q_2}$ (for more information see [18]). The maximum $|\Delta_L(\theta,\varphi)|$ values are 0.08, 0.07 and 0.04 for Ge, Si and diamond crystals respectively; therefore, the terms linear in $\Delta_L(\theta,\varphi)$ alone can be taken into account, whereas the quadratic terms can be neglected. This approximation is accurate to within 1%. The expression for the matrix element describing the three-phonon scattering of quasi-transverse phonons in the long-wave approximation has the form [18]:

$$I(\theta_1,\varphi_1,\theta_2,\varphi_2) = \left(2\frac{q_1 q_2^3}{q_3}\right)^{-2} \left|V_{q_1 q_2 q_3}^{TLL}\right|^2 = \{A_{cub}(\mathbf{e_1 n_2})\cos\theta_{12} + (2\tilde{c}_{155} + \Delta C)N_C + \\ + 2\tilde{c}_{155}N_{155} + 0.5\tilde{c}_{111}N_{111} + \tilde{c}_{112}N_{112} + (\mathbf{e_1 n_1})[0.5\tilde{c}_{112}\Delta N_{112} + 0.5\tilde{c}_{123} + (0.5c_{12} + c_{144})]\}^2, \qquad (10)$$

where $A_{cub} = c_{12} + 3c_{44} + 2c_{144} + 4c_{456}$, $y = \frac{q_1}{q_2} = \frac{z_1}{z_2}\frac{(1+\Delta_2)}{S^*(\theta_1,\varphi_1)}$,

$$N_C = \sum_i n_{1i} n_{2i}^2 [(\mathbf{e_1 n_2})n_{2i} + 0.5e_{1i}], \quad N_{112} = \sum_i e_{1i} n_{1i} n_{2i}^2, \quad N_{111} = \sum_i e_{1i} n_{1i} n_{2i}^4, \qquad (11)$$

$N_{155} = \sum_i e_{1i} n_{2i}^2 [n_{2i}\cos\theta_{12} + 0.5n_{1i}]$, $\Delta N_{112} = \sum_i n_{2i}^2 n_{2i}^2$, $\Delta N_{155} = \sum_i n_{1i} n_{2i}^3$,

$\tilde{c}_{111} = c_{111} - 3c_{112} + 2c_{123} + 12c_{144} - 12c_{155} + 16c_{456}$, $\Delta C = c_{11} - c_{12} - 2c_{44}$,

$\tilde{c}_{112} = c_{112} - c_{123} - 2c_{144}$, $\tilde{c}_{155} = c_{155} - c_{144} - 2c_{456}$,

where **n** = **q**/q = (sin($\theta$)cos($\varphi$), sin($\theta$) sin($\varphi$), cos($\theta$)) is the unit vector of a phonon, $c_{ij}$ denotes the second-order elasticity moduli, and $c_{ijk}$ stands for the third-order elasticity moduli. The first term proportional to $A_{cub}$ in (10) corresponds to the isotropic scattering, while the other terms containing the third-order elasticity moduli $\tilde{c}_{111}$, $\tilde{c}_{112}$, $\tilde{c}_{155}$ and $\Delta C$ correspond to the anisotropic scattering. These terms distinguish cubic crystals from isotropic media: they turn to zero in the isotropic medium model [12]. It should be noted that in the arbitrary direction of the wave vector $\mathbf{q}_1$, unlike the [001] direction considered in [12], the matrix element (10) includes new terms proportional to the third-order elasticity moduli $\tilde{c}_{111}$ and $\tilde{c}_{112}$, as well as the term proportional to $(\mathbf{e}_1\mathbf{n}_1)$, which takes into account the longitudinal component of the transverse-longitudinal vibrations.

According to Matthiessen's rule, the scattering by defects and the normal processes of the phonon scattering in the Landau-Rumer mechanism make an additive contribution to the transverse sound relaxation. Therefore, formulas (1) and (6)-(11) give for the absorption:

$$\alpha_T(\theta_1, \varphi_1, T) = 8.68 \left( B_0^{TLL}(\theta_1, \varphi_1) T_q T^4 + B_i T_q^4 \right) / 2 S_t(\theta_1, \varphi_1) (dB/cм). \tag{12}$$

It is seen from (12) that at $T << T_q$ the coefficient $\alpha_T$ tends to the constant, which is characteristic of the scattering on defects, while at $T >> T_q$ it follows the $T^4$ dependence typical of the Landau-Rumer mechanism. Formulas (6)-(12) can be used to calculate dependences of the ultrasound absorption on the direction of the quasi-transverse wave vector when the cubic anisotropy is taken into account exactly in the energy conservation law subject to the competition between the anharmonic scattering processes and the scattering by defects. For analysis of these dependences, we shall introduce the dimensionless coefficient $\alpha_T^*(\theta_1, \varphi_1)$ characterizing the change of the ultrasound absorption relative to the [001] direction:

$$\alpha_T^*(\theta_1, \varphi_1) = \frac{\alpha_T(\theta_1, \varphi_1)}{\alpha_{T[100]}(0,0,T)} = \frac{S_{[100]}^t}{S_t(\theta_1, \varphi_1)} \frac{\left(B_0^{TLL}(\theta_1, \varphi_1) T^4 + B_i T_q^3\right)}{\left(B_{0[100]}^{TLL} T^4 + B_i T_q^3\right)} = \left(\frac{S_{[100]}^t}{S_t(\theta_1, \varphi_1)}\right)^2 \frac{\left(J(\theta_1, \varphi_1) T^4 + B_i T_q^3 S_t(\theta_1, \varphi_1)\right)}{\left(J_{[100]} T^4 + B_i T_q^3 S_{[100]}^t\right)}. \tag{13}$$

The ultrasound absorption in the [001] direction is defined as

$$\alpha_{T[100]} = \left(A_{[100]}^{TLL} T_q T^4 + A_{i\,[100]} T_q^4\right) (dB/cм), \quad A_{[100]}^{TLL} = \frac{4.34}{S_{[100]}^t} B_{0[100]}^{TLL}, \quad A_{i\,[100]} = \frac{4.34}{S_{[100]}^t} B_i. \tag{14}$$

It follows from (13) that for the scattering by defects the coefficient $\alpha_i^*(\theta_1, \varphi_1)$ has the form

$$\alpha_i^*(\theta_1, \varphi_1) = S_{[100]}^t / S_t(\theta_1, \varphi_1). \tag{15}$$

In the case of the Landau-Rumer mechanism, the ultrasound absorption anisotropy $\alpha_{TLL}^*(\theta_1, \varphi_1)$ is defined by the expression

$$\alpha^*_{TLL}(\theta_1,\varphi_1) = \frac{S^t_{[100]}}{S_t(\theta_1,\varphi_1)} \frac{B^{TLL}_0(\theta_1,\varphi_1)}{B^{TLL}_{0[100]}} = \left(\frac{S^t_{[100]}}{S_t(\theta_1,\varphi_1)}\right)^2 \frac{J(\theta_1,\varphi_1)}{J_{[100]}}. \tag{16}$$

We shall show below that the ultrasound absorption anisotropies are considerably different for the aforementioned two cases.

### 3. Results of Numerical Analysis of Transverse Ultrasound Absorption in Cubic Crystals

Let us analyze the angular dependences of the acoustic wave absorption for two most important cases: (1) the acoustic wave vectors $q_1$ are in the cube face plane ($\varphi_1 = 0$); (2) the acoustic wave vectors $q_1$ are in the diagonal plane ($\varphi_1 = \pi/4$). These dependences are shown in Figs. 1 to 5 for cubic crystals with the positive (Ge, Si, diamond, GaSb and InSb) and the negative (KCl, NaCl and GaAs) anisotropy of the second-order elasticity moduli. The calculations were made using experimental values of the thermodynamic elasticity moduli of the second $c_{ik}$ and third $c_{ijk}$ orders, which were adopted from [1, 23] (see Table 1).

**Table 1.** Thermodynamic elasticity moduli for the cubic crystals under study, in $10^{12}$ dyne/cm$^2$. The data are adopted from [1, 23]

|  | Ge | Si | Алмаз | InSb | GaSb | KCl | NaCl | GaAs |
|---|---|---|---|---|---|---|---|---|
| $c_{11}$ | 1.289 | 1.657 | 10.76 | 0.672 | 0.885 | 0.398 | 0.487 | 0.192 |
| $c_{12}$ | 0.483 | 0.638 | 1.25 | 0.367 | 0.404 | 0.062 | 0.124 | 0.0599 |
| $c_{44}$ | 0.671 | 0.796 | 5.758 | 0.302 | 0.433 | 0.0625 | 0.126 | 0.0538 |
| $\Delta C$ | 0.54 | 0.57 | 2.01 | 0.3 | 0.385 | -0.211 | -0.11 | -0.0245 |
| $c_{111}$ | -7.10 | -8.25 | -62.6 | -3.56 | -4.75 | -7.01 | -8.8 | -6.22 |
| $c_{112}$ | -3.89 | -4.51 | -22.6 | -2.66 | -3.08 | -0.571 | -0.571 | -3.87 |
| $c_{123}$ | -0.18 | -0.64 | 1.12 | -1.0 | -0.44 | 0.284 | 0.284 | -0.57 |
| $c_{144}$ | -0.23 | 0.12 | -6.74 | 0.16 | 0.5 | 0.127 | 0.257 | 0.02 |
| $c_{155}$ | -2.92 | -3.10 | -28.6 | -1.39 | -2.16 | -0.245 | -0.611 | -2.69 |
| $c_{456}$ | -0.53 | -0.64 | -8.23 | -0.004 | -0.25 | 0.118 | 0.271 | -0.39 |
| $A_{cub}$ | -0.084 | 0.71 | -27.88 | 1.57 | 1.7 | 0.98 | 2.09 | -1,29 |
| $\widetilde{c}_{155}$ | -1.63 | -1.9 | -5,4 | -1.54 | -2.16 | -0.61 | -1.41 | -1.93 |
| $\widetilde{c}_{111}$ | 28.01 | 32.4 | 138.1 | 20.96 | 31.53 | 1.62 | 8.23 | 30.53 |
| $\widetilde{c}_{112}$ | -3.25 | -4.1 | -10.24 | -1.98 | -3.64 | -1.11 | -1.37 | -3.34 |

It should be noted first that in the case of the scattering by defects, the ultrasound absorption anisotropy is determined by the harmonic energy anisotropy. The angular dependences of the transverse ultrasound absorption in the cubic crystals of the first and second types are qualitatively different (see Fig. 1). For example, for a quasi-transverse mode $t_2$, whose wave and polarization vectors are in the cube face plane ($\varphi_1 = 0$), the absorption $\alpha_{i\,t2}^{*}(\theta_1,0)$ is maximum and minimum in the [101] and [001] crystallographic directions respectively in the Ge, Si, diamond, InSb and GaSb crystals. The absorption $\alpha_{i\,t2}^{*}(\pi/4,0)$ is equal to 1.29, 1.25, 1.1, 1.41 and 1.34 in the Ge, Si, diamond, InSb and GaSb crystals, respectively. In the crystals of the second type (KCl, NaCl and GaAs), these dependences exhibit the inverse behavior (the absorption maximums and minimums exchange places): the absorption $\alpha_{i\,t2}^{*}(\theta_1,0)$ is maximum in directions like [101] and minimum in directions like [001]. The absorption $\alpha_{i\,t2}^{*}(\pi/4,0)$ is equal to 0.61, 0.83 and 0.9 in KCl, NaCl and GaAs, respectively

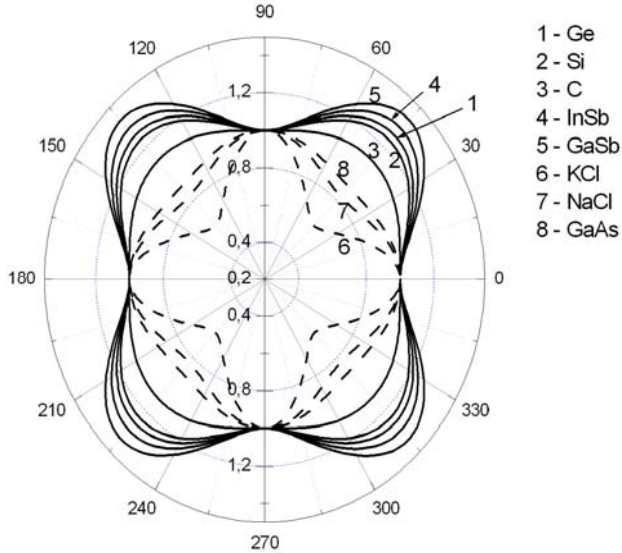

**Fig. 1.** Dependences of the reduced ultrasound absorption $\alpha_i^{*}(\theta_1,\varphi_1)$ on the wave vector direction for the quasi-transverse mode $t_2$ in the case of the scattering by defects.

If the anharmonic processes play the dominant role, the ultrasound absorption anisotropy is determined by both the harmonic energy anisotropy and the anharmonic energy anisotropy in the crystals. In this case, the angular dependences of the absorption $\alpha_{TLL\,t2}^{*}(\theta_1,\varphi_1)$ are inverse to those for the scattering by defects. In the crystals of the first type (Ge, Si, diamond, InSb and GaSb), the absorption $\alpha_{TLL\,t2}^{*}(\theta_1,\varphi_1)$ is maximum in crystallographic directions like [100] (Ge, Si and diamond) and directions near [100] (InSb and GaSb), while the minimum is observed in directions like [101] and [111]. On the contrary, in the crystals of the second type (KCl, NaCl), the absorption $\alpha_{TLL\,t2}^{*}(\theta_1,\varphi_1)$ is maximum in crystallographic directions like [101] and [111], and minimum in directions like [001]. The maximum absorption in the ionic KCl and NaCl crystals proves to be two orders of magnitude larger than the maximum absorption in Ge, whereas these values are similar with respect to their order of magnitude in crystallographic directions like [100]. Thus, the angular dependences of the absorption are qualitatively different for the crystals of the first and second types. The absorption decreases in the crystals of the first type by nearly one order of magnitude in going from Ge to Si and Si to diamond. This decrease is explained mainly by the change of the coefficient $B_0$ (see formula (7)), which depends on the second-order elasticity moduli (see Table 1). The coefficient $B_0$ decreases by one order of magnitude in going from Ge to Si and by three orders of magnitude in going from Si to diamond. However, the two orders of magnitude are compensated in the diamond crystals by large values of the third-order elasticity moduli, which determine the probability of the anharmonic scattering processes (see Table 1). The growth of the absorption by two orders of magnitude in KCl and four orders of magnitude in GaAs as compared with the absorption in KCl is due mainly to the change of the parameter $B_0$, which depends on the second-order elasticity moduli (see Table 2). Notice that the second-order elasticity moduli $c_{ik}$ are anomalously small in GaAs. They are one order of magnitude smaller than the moduli in the Ge and InSb crystals (see Table 1).

**Table 2.** Parameters and the transverse ultrasound absorption in the cubic crystals under study

|  | $B_0\ 10^{24}$ (cm$^4$ dyn$^{-2}$ s$^{-1}$K$^{-5}$) [100] | $B_0^{TLL}$ (s$^{-1}$K$^{-5}$) [100] | $\alpha_{TLL[001]}$, (dB/sm) | $\alpha^*_{TLL\ t2}$ [101] | $\alpha^*_{TLLt\ 1}$ [101] | $\alpha^*_{TLL\ t1}$ [111] | $\alpha^*_{TLL\ t2}$ [111] | $\Phi,\ 10^{-16}$ (s/sm)$^3$ |
|---|---|---|---|---|---|---|---|---|
| Ge | 0.239 | 0.87 | 1.06 | 0.23 | 0.64 | 0.32 | 0.32 | 2.88 |
| Si | 0.026 | 0.071 | 5.27·10$^{-2}$ | 0.33 | 0.73 | 0.37 | 0.37 | 0.615 |
| Алмаз | 1.3·10$^{-5}$ | 0.0057 | 1.93·10$^{-3}$ | 0.11 | 0.798 | 0.31 | 0.31 | 0.052 |
| InSb | 5.23 | 2.43 | 4.61 | 0.47 | 0.86 | 0.39 | 0.39 | 11.66 |
| GaSb | 1.418 | 1.42 | 2.22 | 0.41 | 0.7 | 0.36 | 0.36 | 6.25 |
| KCl | 88.67 | 0.627 | 1.54 | 107 | 0.48 | 85 | 85 | 9.47 |
| NaCl | 19.32 | 1.13 | 2.03 | 37 | 1.37 | 25 | 25 | 5.28 |
| GaAs | 3.99·10$^3$ | 1,63·10$^4$ | 7·10$^4$ | 0.29 | 1.26 | 0.46 | 0.46 | 79.44 |

Let us consider in more detail on the dominant role of the anharmonic scattering processes for the acoustic wave vectors $\mathbf{q}_1$ in the cube face plane ($\varphi_l = 0$). The isotropic mode $t_1$ ($S_{t1}(\theta_1, 0) = \sqrt{c_{44}/\rho} = const$), whose polarization vector is perpendicular to the cube face at hand, is a fast mode in the crystals of the first type and a slow mode in the crystals of the second type. In the case of the scattering by defects, the absorption $\alpha^*_{i\ t1}(\theta_1, 0)$ of this mode is isotropic in the crystals of both the first and second types. In terms of the Landau-Rumer mechanisms, the absorption $\alpha^*_{TLL\ t1}(\theta_1, 0)$ changes little with the angle $\theta_l$ in the Ge, Si, diamond, InSb and GaSb crystals (see Fig. 2a, curves 1 to 5). The absorption $\alpha^*_{TLL\ t1}(\theta_1, 0)$ reaches its minimum in [101] directions ($\theta_l = \pi/4$) in compounds of the first group. The $\alpha^{*}_{TLL\,t1}(\pi/4,0)$ values are 0.64, 0.73, 0.8, 0.85 and 0.7 in Ge, Si, diamond, InSb and GaSb, respectively. The $\alpha^{*}_{TLL\,t1}(\theta_1,0)$ values are maximum in [001] directions in the Ge and diamond crystals. Unlike in Ge and diamond, local minimums are observed in [001] directions in InSb, GaSb and Si. The absorption $\alpha^{*}_{TLL\,t1}(\theta_1,0)$ is maximum at the angles $\theta_l = n\pi/2 \pm 0.3$, $\theta_l = n\pi/2 \pm 0.19$ and $\theta_l = n\pi/2 \pm 0.13$ (n = 0, 1, 2, etc.) and is equal to 1.18, 1.05 and 1.002 in the InSb, GaSb and Si crystals, respectively (see Fig. 2a, curves 2, 4 and 5). The ratio between the minimum and maximum values is 0.73, 0.67 and 0.73 in InSb, GaSb and Si, respectively. These specific features of the absorption in the InSb, GaSb and Si crystals are due mainly to the anharmonic energy anisotropy. The point is that in the case of the pure mode $t_1$ with $\mathbf{n}_1 = \{\sin\theta_1, 0, \cos\theta_1\}$ and the polarization vector $\mathbf{e}_1 = \{0,1,0\}$, $(\mathbf{e}_1\mathbf{n}_1) = 0$ and $e_{1i}n_{1i} = 0$ at all $i$, while $N_{111} = N_{112} = 0$. Therefore, the matrix element (10) only contains the terms proportional to the third-order elasticity moduli $A_{cub}$ and $\tilde{c}_{155}$:

$$I(\theta_1,\varphi_1,\theta_2,\varphi_2) = n_{2y}^2\left[\left(A_{cub} + 2\tilde{c}_{155}n_{2y}^2\right)\cos\theta_{12} + \left(2\tilde{c}_{155} + \Delta C\right)\left(n_{1x}n_{2x}^3 + n_{1z}n_{2z}^3\right)\right]^2. \quad (17)$$

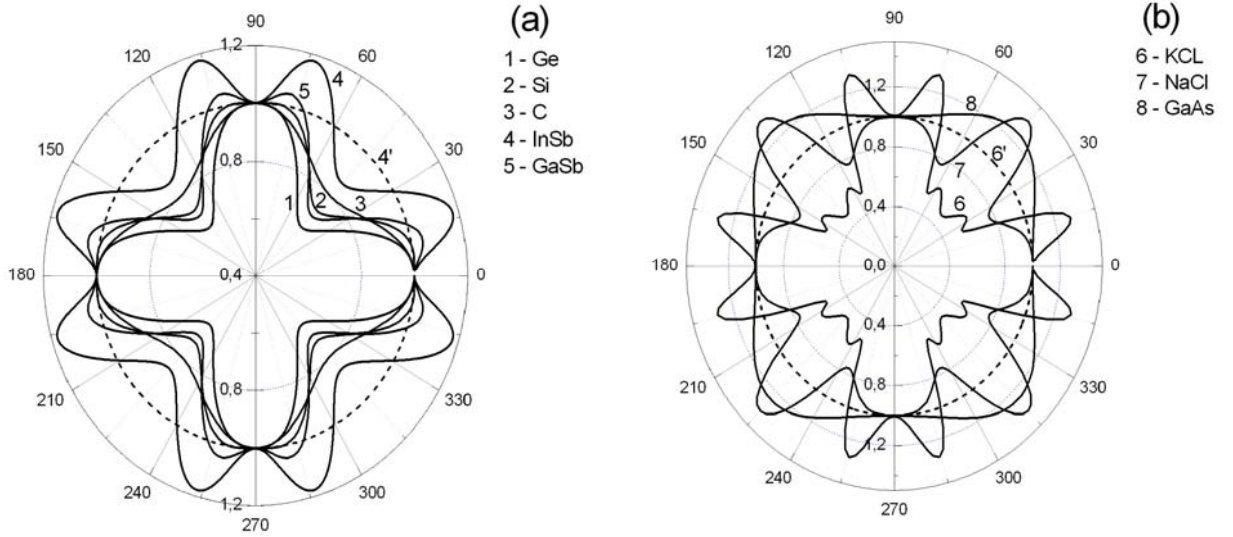

**Fig. 2.** Angular dependences of the absorption $\alpha^{*}_{TLL\,t1}(\theta_1,0)$ for the pure mode $t_1$ with the wave vector in the cube face plane and the polarization vector perpendicular to this plane in the case of the Landau-Rumer mechanism: (a) crystals of the first type; (b) crystals of the second type. Dashed curves 4', 6': scattering by defects.

The moduli $A_{cub}$ and $\tilde{c}_{155}$ have opposite signs in these crystals (see Table 1). Therefore, the contributions of the isotropic ($A_{cub}$) and anisotropic ($\tilde{c}_{155}$) scatterings to the absorption are mutually compensated to a large extent, unlike in Ge and diamond, where this compensation is absent. Since the compensation is larger in InSb than in GaSb and Si, the dependence $\alpha^*_{TLL\,t1}(\theta_1,0)$ has a deeper minimum in directions like [100]. Thus, the local minimums in the dependences $\alpha^*_{TLL\,t1}(\theta_1,0)$ for the [100] directions in the InSb, GaSb and Si crystals appear due to the mutual compensation of the terms proportional to the elasticity moduli $A_{cub}$ and $\tilde{c}_{155}$ in the matrix element of the three-phonon scattering processes.

For the Landau-Rumer mechanism, the angular dependences of the ultrasound absorption, $\alpha^*_{TLL\,t1}(\theta_1,0)$, are more complicated in the KCl and NaCl crystals of the second type than they are in the crystals of the first type (see Fig. 2b). The absorption $\alpha^*_{TLL\,t1}(\theta_1,0)$ has the maximum value of 1.32 in the KCl crystal at the angle $\theta_l \cong n\pi/2\pm0.25$ and the minimum value of $\alpha^*_{TLL\,t1}(\pi/4,0) = 0.48$ at the angle $\theta_l = \pi/4$. At the angles $\theta_l = 0$, $\theta_l \cong 0.44$, $\theta_l \cong 1.13$ and $\theta_l = \pi/2$ the function $\alpha^*_{TLL\,t1}(\theta_1,0)$ has four local minimums. The maximum values of $\alpha^*_{TLL\,t1}(\theta_1,0)$ at the angles $\theta_l \cong n\pi/2\pm0.25$ are nearly three times as large as the minimum values of $\alpha^*_{TLL\,t1}(\pi/4,0)$. In the NaCl crystal the maximum absorption $\alpha^*_{TLL\,t1}(\theta_1,0)$ equal to 1.37 is observed in directions like [101] ($\theta_l = \pi/4$), while the minimum $\alpha^*_{TLL\,t1}(\theta_1,0) = 0.76$ is achieved at the angles $\theta_l \cong 0.44$ and $\theta_l \cong 1.13$. In directions like [001] ($\theta_l = 0$, $\pi/2$) the function $\alpha^*_{TLL\,t1}(\theta_1,0)$ has two local maximums. In the GaAs crystal the angular dependences of the absorption $\alpha^*_{TLL\,t1}(\theta_1,0)$ are inverse to those in the crystals of the first group: the maximum values are achieved in [101] directions ($\theta_l = \pi/4$), while the minimum values in [001] directions (see Fig. 2b, curve 8). The angular dependences $\alpha^*_{TLL\,t1}(\theta_1,0)$ are different in the crystals of the second group because the elasticity moduli $A_{cub}$ and $\tilde{c}_{155}$ have opposite signs in the KCl and NaCl crystals and their contributions to the absorption are mutually compensated to a greater (KCl) or lesser (NaCl) extent. These moduli have like signs in the GaAs crystals and such compensation is absent.

In the case of the quasi-transverse mode $t_2$ with the wave and polarization vectors in the cube face plane and the anharmonic scattering processes playing the dominant role, the absorption $\alpha^*_{TLL\,t2}(\theta_1,0)$ in the crystals of the first and second types changes to a greater degree than in the case of the pure mode $t_1$ (see Figs. 3a and 3b). In the crystals with the positive anisotropy of the second-order elasticity moduli, the absorption $\alpha^*_{TLL\,t2}(\theta_1,0)$ is minimum in [101] directions ($\theta_l = \pi/4$). Here $\alpha^*_{TLL\,t2}(\pi/4,0)$ equals 0.23, 0.33, 0.11, 0.47 and 0.4 in Ge, Si, diamond, InSb and GaSb, respectively. The maximum $\alpha^*_{TLL\,t2}(\theta_1,0)$ values are achieved in [001] directions in the Ge, Si and diamond crystals, whereas local minimums are observed in these directions in InSb and GaSb. The maximum values of the absorption $\alpha^*_{TLL\,t2}(\theta_1,0)$ are achieved at the angles $\theta_l \cong n\pi/2\pm0.19$ and $n\pi/2\pm0.1$ (n = 0, 1, 2, etc.) and are equal to 1.3 and 1.02 in InSb and GaSb respectively (see Fig. 2a, curves 4 and 5). The ratio of the minimum to maximum values of the absorption $\alpha^*_{TLL\,t2}(\theta_1,0)$ is 0.36 and 0.39 in InSb and GaSb, respectively. As distinct from the pure mode $t_1$ discussed above, all the terms of the matrix element (10) contribute to the absorption of the quasi-transverse mode $t_2$. The symmetric maximums, which appear in the dependences $\alpha^*_{TLL\,t2}(\theta_1,0)$ in directions like [001] in the InSb and GaSb crystals, are due to both the anisotropy of the anharmonic and the harmonic energy of the crystals. The anisotropy of the harmonic energy of the crystals gives rise to the longitudinal component of the quasi-transverse mode $t_2$, which plays a dominant role in the formation of the symmetric maximums. Figure 3a (dashed line 4a) presents the dependence of the mode $t_2$ absorption in InSb as an approximation of the pure mode, i.e. when the polarization vector of the quasi-transverse mode is replaced by the polarization vector of the pure mode ($e_0^{t2}$ = (cos$\theta_1$, 0, -sin$\theta_1$)). In this case, the local minimums near directions like [001] vanish for the pure mode $t_2$ in the InSb and GaSb crystals. It should be noted that the anisotropy of the anharmonic energy of InSb and GaSb is also important for the realization of this specific feature. The point is that in Ge the maximum contribution of the longitudinal component to the mode $t_2$ is 15.5%, which differs little from 15% in InSb and GaSb. However, the dependence $\alpha^*_{TLL\,t2}(\theta_1,0)$ exhibits a clear-cut maximum in [100] direction in the Ge crystal. As indicated above, the absorption of the ultrasound propagating along the cube edges is contributed only by the terms proportional to the third-order elasticity moduli $A_{cub}$ and $\tilde{c}_{155}$ (see formula (17)), which have opposite signs in the InSb and GaSb crystals (see Table 1). Consequently, their contributions are mutually compensated to a great extent and, therefore, the ultrasound absorption in these directions is considerably smaller than the absorption in Ge where this compensation is absent. Since the compensation is more complete in InSb than in GaSb, a deeper minimum is formed in the dependences $\alpha^*_{TLL\,t2}(\theta_1,0)$. Thus, the appearance of local maximums in $\alpha^*_{TLL\,t2}(\theta_1,0)$ near [100] directions in the InSb and GaSb crystals is due to both the longitudinal component of the quasi-transverse mode $t_2$ and the mutual compensation of the terms proportional to the third-order elasticity moduli $A_{cub}$ and $\tilde{c}_{155}$ in the matrix element of the three-phonon scattering processes. According to relevant estimates, the maximum contribution of the longitudinal component to the absorption of quasi-transverse modes is 50% in InSb, 34% in GaSb, 30% in Ge and Si, and 6% in diamond. The comparison of Figs. 1 and 2a shows that in the case of the scattering by defects, the angular dependences of the absorption, $\alpha^*_{i\,t2}(\theta_1,0)$, in the crystals of the first type are inverse to the corresponding dependences observed in the case of the anharmonic scattering processes.

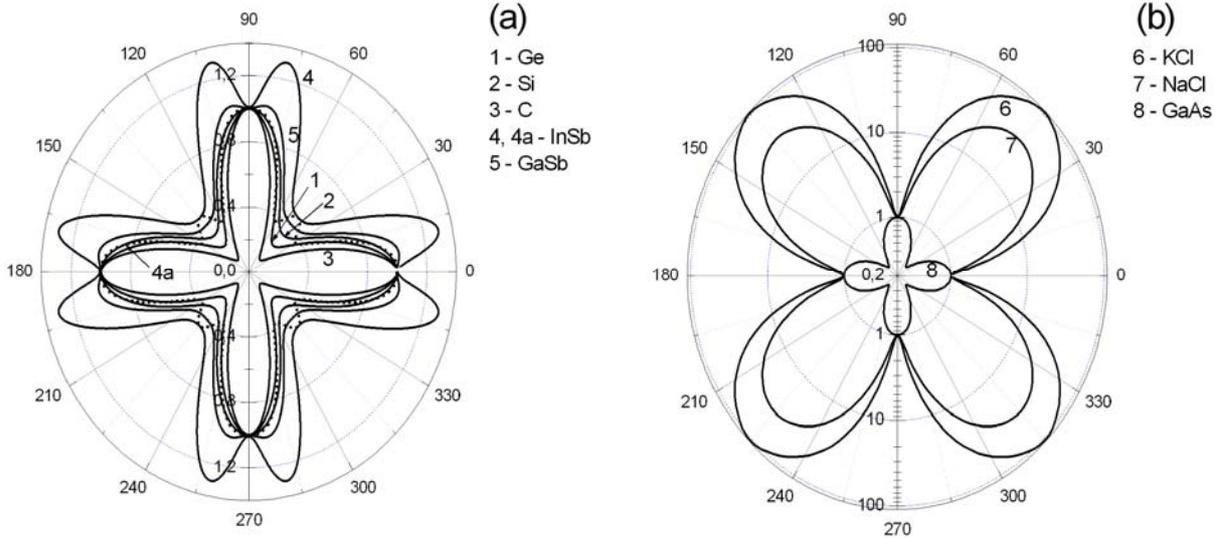

**Fig. 3.** Angular dependences of the absorption $\alpha^*_{TLL\,t2}(\theta_1,0)$ for the quasi-transverse mode $t_2$ with the wave and the polarization vector in the cube face plane for the case of the Landau-Rumer mechanism: (a) crystals of the first type; (b) crystals of the second type. Curve 4a: InSb in the pure mode approximation.

For the quasi-transverse mode $t_2$, the angular dependences of the ultrasound absorption in the crystals of the second type (KCl and NaCl) are inverse to the corresponding dependences in the crystals of the first type (see Fig. 3b, curves 6 and 7). The absorption $\alpha^*_{TLL\,t2}(\theta_1,0)$ in KCl and NaCl increases sharply from the minimum at $\theta_1 = 0$, reaches the maximum at $\theta_1 = \pi/4$, and decreases to the initial value at $\theta_1 = \pi/2$. In the GaAs crystal, unlike in KCl and NaCl, the angular dependence of the absorption, $\alpha^*_{TLL\,t2}(\theta_1,0)$, is the same as the corresponding dependence in Ge, Si and diamond (see Fig. 3a, curves 1 to 3). The dependence $\alpha^*_{TLL\,t2}(\theta_1,0)$ includes maximums in [001] and [100] directions, and minimums in [101] direction. The absorption $\alpha^*_{TLL\,t2}(\pi/4,0)$ is 107, 37 and 0.29 in KCl, NaCl and GaAs, respectively. The maximum contribution of the longitudinal component to the absorption of the mode $t_2$ is 22% in KCl and 3% in NaCl and GaAs. The comparison of Figs. 1 and 3b demonstrates that the angular dependences of the ultrasound absorption in KCl and NaCl observed for the scattering by defects are inverse to those in the case of the anharmonic scattering processes: the positions of the maximums and the minimums are reversed.

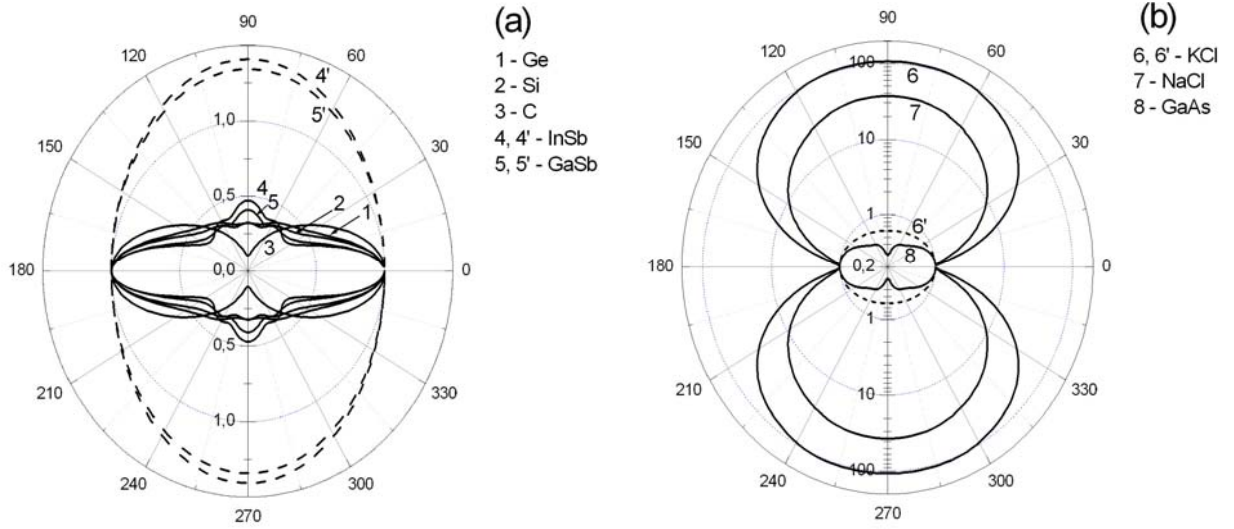

**Fig. 4.** Dependences of the absorption $\alpha^*_{TLL\,t1}(\theta_1,\pi/4)$ as a function of the wave-vector for the pure mode $t_1$ with the polarization vector perpendicular to the diagonal plane in the case of the Landau-Rumer mechanism: (a) crystals of the first type; (b) crystals of the second type. Dashed curves 4', 5' and 6': scattering by defects in the InSb, GaSb and KCl crystals, respectively.

For the pure mode $t_1$ with the wave vector in the diagonal plane ($\varphi_l = \pi/4$) and the polarization vector perpendicular to this plane, the absorption $\alpha^*_{TLL\,t1}(\theta_1,\pi/4)$ is maximum in the [001] direction in Ge, Si, diamond, InSb and GaSb (see Fig. 4a, curves 1 to 5). As the angle $\theta_l$ decreases, the absorption diminishes to its minimum value in the [110] direction ($\theta_l = \pi/2$) in the Ge, Si and diamond crystals and equals 0.32, 0.33 and 0.11, respectively (see Fig. 4a, curves 1 to 3). In InSb and GaSb the absorption reaches its minimum values at angles near $\theta_l \cong \pi/4$, while a local maximum appears in the [110] direction in these crystals (see Fig. 4a, curves 4 and 5). Notice that the relationship $\alpha^*_{TLL\,t1}(\pi/2,\pi/4)=\alpha^*_{TLL\,t2}(\pi/4,0)$ holds for the absorption referring to different modes. The situation is reverse in the crystals of the second group (KCl and NaCl): the absorption $\alpha^*_{TLL\,t1}(\theta_1,\pi/4)$ is minimum in the [001] direction ($\theta_l = 0$), increases monotonically with the angle $\theta_l$, and becomes maximum in the [110] direction (see Fig. 4b, curves 6 and 7). The angular dependence of the absorption $\alpha^*_{TLL\,t1}(\theta_1,\pi/4)$ in GaAs is inverse to the corresponding dependence in KCl and NaCl. It is analogous to the dependence $\alpha^*_{TLL\,t1}(\theta_1,\pi/4)$ in the crystals of the first group (Ge, Si and diamond) (see Fig. 4a, curves 1 to 3). The absorption $\alpha^*_{TLL\,t1}(\theta_1,0)$ is maximum in the [001] direction and minimum in the [110] direction (see Fig. 4b, curve 8). The absorption $\alpha^*_{TLL\,t1}(\pi/2,\pi/4)$ is 107, 37 and 0.29 in KCl, NaCl and GaAs, respectively. In the case of the scattering by defects, in contrast to the anharmonic scattering processes, the absorption becomes maximum in the [110] direction and minimum in the [001] direction in the crystals of the first group (see Fig. 4a, curves 4' and 5'). The situation is reverse in the crystals of the second group. The absorption $\alpha^*_{i\,t2}(\pi/2,\pi/4)$ equals 1.20, 1.17, 1.07, 1.26, 1.23, 0.68, 0.87 and 0.93 in Ge, Si, diamond, InSb, GaSb, KCl, NaCl and GaAs, respectively.

For the quasi-transverse mode $t_2$ with the polarization vector in the diagonal plane ($\varphi_l = \pi/4$), the absorption $\alpha^*_{TLL\,t2}(\theta_1,\pi/4)$ in the Ge, Si and diamond crystals is also considerably different from the absorption in the InSb and GaSb crystals (see Fig. 5a, curves 1 to 5). The absorption $\alpha^*_{TLL\,t2}(\theta_1,\pi/4)$ has minimum values in the crystals of the first group at angles $\theta_l \cong \pi/4$ and equals 0.27, 0.33, 0.23, 0.38 and 0.34 in Ge, Si, diamond, InSb and GaSb, respectively. In Ge, Si and diamond the absorption $\alpha^*_{TLL\,t2}(\theta_1,\pi/4)$ decreases from the maximum value in the [001] direction ($\theta_l = 0$), becomes minimum at angles $\theta_l \approx \pi/4$, increases again, and then at $\theta_l = \pi/2$ approaches the value of $\alpha^*_{TLL\,t1}(\pi/4,0)$, which corresponds to the fast mode in the cube face plane at $\theta_l = \pi/4$ (see Fig. 5a, curves 1 to 3). Local minimums appear in the [100] direction in InSb and GaSb, unlike in the Ge, Si and diamond crystals. The absorption $\alpha^*_{TLL\,t2}(\theta_1,\pi/4)$ is maximum at angles $\theta_l = n\pi/2\pm0.22$ and $n\pi/2\pm0.16$ and equals 1.6 and 1.15 in InSb and GaSb, respectively (see Fig. 5a, curves 4 and 5). The ratio of the minimum to maximum values of the absorption $\alpha^*_{TLL\,t2}(\theta_1,\pi/4)$ is 0.36 and 0.39 in InSb and GaSb, respectively. As noted above, local maximums in the dependence $\alpha^*_{TLL\,t2}(\theta_1,\pi/4)$ near the [100] direction in the InSb and GaSb crystals appear due to both the longitudinal component of the quasi-transverse mode $t_2$ and the mutual compensation of the terms proportional to the third-order elasticity moduli $A_{cub}$ and $\tilde{c}_{155}$ in the matrix element of the three-phonon scattering processes. The calculation of the absorption of the quasi-transverse mode $t_2$ in the approximation of the pure mode (the polarization vector of the mode $t_2$ is replaced by the polarization vector of the pure mode $e_0^{t2} = (\cos\theta/\sqrt{2}, \cos\theta/\sqrt{2}, -\sin\theta)$) shows that in the case of the pure mode $t_2$ the local maximums considerably diminish near directions like [001] in InSb and GaSb and become as low as 1.14 and 1.01, respectively (see Fig. 5a, curve 4a).

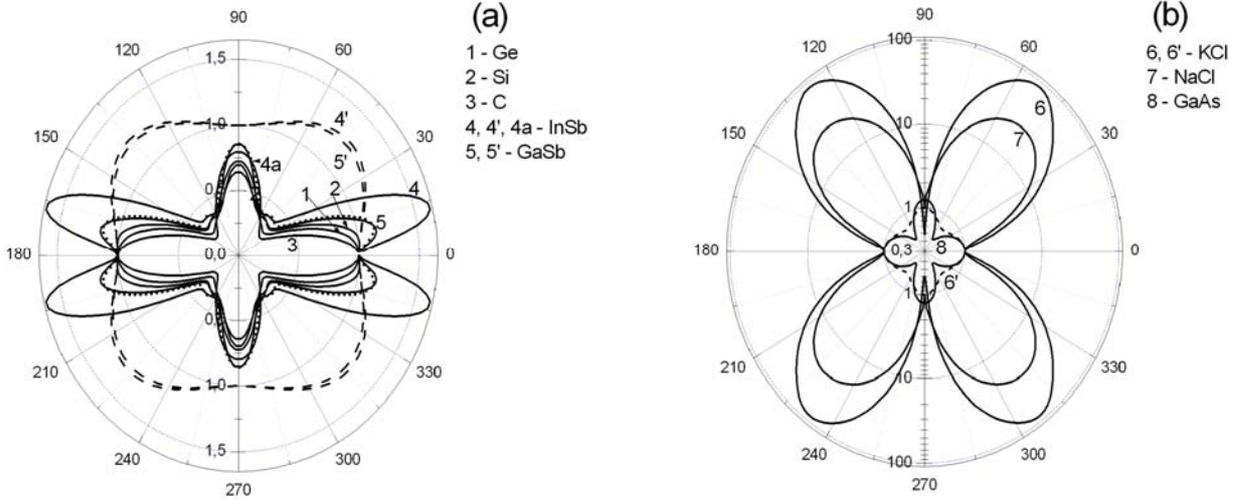

**Fig. 5.** Angular dependences of the absorption $\alpha^*_{TLL\,t2}(\theta_1, \pi/4)$ for the quasi-transverse mode $t_2$ with the wave and the polarization vector in the diagonal plane in the case of the Landau-Rumer mechanism: (a) crystals of the first type; (b) crystals of the second type. Curve 4a: InSb in the pure mode approximation. Dashed curves 4', 5' and 6': scattering by defects in the InSb, GaSb and KCl crystals, respectively

According to relevant estimates, the maximum contribution of the longitudinal component to the absorption $\alpha^*_{TLL\,t2}(\theta_1, \pi/4)$ of the quasi-transverse modes is 35% in InSb, 20% in Ge, Si and GaSb, and 11% in diamond. If the scattering by defects is concerned, the angular dependences of the absorption change qualitatively (see Fig. 5a, curves 4' and 5'): the absorption $\alpha^*_{i\,t2}(\theta_1, \pi/4)$ is minimum in the [001] and [110] directions and is maximum at angles $\theta_l = \pi/4$. The absorption $\alpha^*_{i\,t2}(\pi/4, \pi/4)$ equals 1.20, 1.17, 1.07, 1.26 and 1.23 for Ge, Si, diamond, InSb and GaSb, respectively.

For the same mode, the angular dependences of the absorption $\alpha^*_{TLL\,t2}(\theta_1, \pi/4)$ in KCl and NaCl and the crystals of the first group differ qualitatively for both the scattering by defects and the Landau-Rumer relaxation mechanism. In the case of the scattering by defects, the absorption $\alpha^*_{i\,t2}(\theta_1, \pi/4)$ peaks in directions like [001] and becomes minimum at angles $\theta_l = \pi/4$ in the KCl, NaCl and GaAs crystals (see Fig. 5b, curve 6'). The absorption $\alpha^*_{i\,t2}(\pi/4, \pi/4)$ equals 0.67, 0.87 and 0.93 in KCl, NaCl and GaAs, respectively. In the case of the anharmonic scattering processes, the absorption $\alpha^*_{TLL\,t2}(\theta_1, \pi/4)$ reaches its maximum values in directions like [111] and its minimum values in the [001] and [011] directions in the KCl and NaCl crystals. The maximum absorption $\alpha^*_{TLL\,t2}(\theta_{[111]},\pi/4)$ is 87 and 27 in KCl and NaCl respectively. As distinct from the KCl and NaCl crystals, in GaAs the absorption is maximum in the [011] direction ($\theta_l = \pi/2$) and minimum in the [111] direction (see Fig. 5b, curve 8), with $\alpha^*_{TLL\,t2}(\theta_{[111]},\pi/4) = 0.46$ and $\alpha^*_{TLL\,t2}(\pi/2,\pi/4) = 1.25$. It should be noted that in KCl the absorption $\alpha^*_{TLL\,t2}(\theta_{[111]},\pi/4)$ proves to be a factor of 1.2 smaller than $\alpha^*_{TLL\,t2}(\pi/4,0)$, while the minimum absorption $\alpha^*_{TLL\,t2}(\pi/2,\pi/4)$ for this mode is 2.1 times larger than $\alpha^*_{TLL\,t2}(0,\pi/4)$. Therefore, the ultrasound absorption for this mode in KCl diminishes 182 times on transition from the [111] direction to [110]. Thus, the anisotropy of the absorption for the quasi-transverse mode $t_2$ is one order of magnitude larger in NaCl and two orders of magnitude larger in KCl than it is in the Ge, Si and diamond crystals. The maximum contribution of the longitudinal component to the absorption of the mode $t_2$ equals 50%, 8% and 7% in KCl, NaCl and GaAs, respectively.

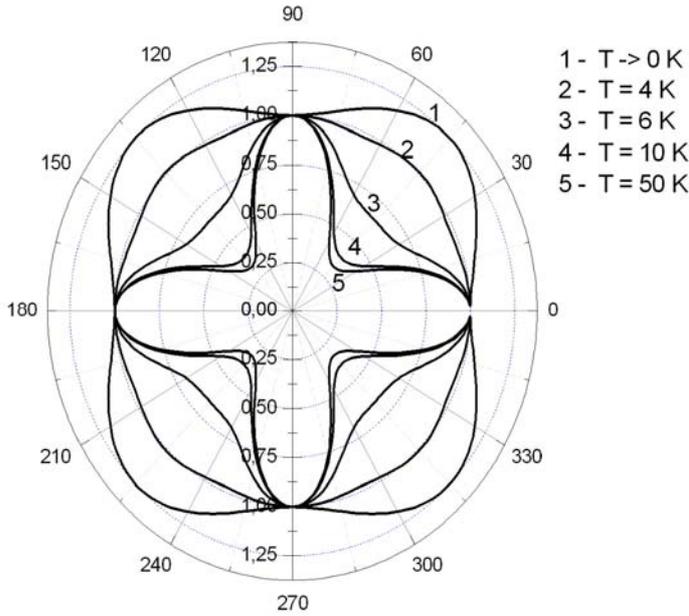

**Fig. 6.** Angular dependences of the absorption for the quasi-transverse mode $t_2$ with the wave and the polarization vector in the cube face plane in the $^{nat}$Si ($g = 2.01\times10^{-4}$) crystals in the presence of the competition between isotopic and anharmonic scattering processes for a fixed energy of ultrasonic quantum $T_q = 4.8$ K and different temperatures.

Let us consider how the angular dependences of the ultrasound absorption $\alpha^*_{T\,t2}(\theta_1,0)$ change with the temperature in the presence of the competition between isotopic and anharmonic scattering processes. We shall take $^{nat}$Si crystals with the natural isotopic composition ($g = 2.01\times10^{-4}$) as an example. It is seen from Fig. 6 (curves 1 to 5) that as the temperature rises from 4 K to 50 K in $^{nat}$Si crystals, the isotopic scattering lets the anharmonic scattering processes have the dominant role and the angular dependences of the absorption $\alpha^*_{T\,t2}(\theta_1,0)$ change qualitatively. The change is most abrupt over the temperature interval of 4 K < $T$ < 10 K when $T_q = 4.8$ K (corresponding to the ultrasound frequency of 100 GHz) is comparable with the temperature. Curve 5 in this figure, which was calculated at T = 50 K, practically coincides with the results obtained for the Landau-Rumer mechanism. It should be noted that in Si crystals enriched to 99.983% in the $^{28}$Si isotope ($g = 3.2\times10^{-7}$) the anharmonic scattering processes make the dominant contribution to the ultrasound absorption at $T_q = 4.8$ K over the temperature interval of 4 K to 50 K, while the angular dependences of the absorption $\alpha^*_{Tt2}(\theta_1,0)$ are described by curve 5. Obviously, from the analysis of the ultrasound absorption anisotropy in cubic crystals it is possible to determine the dominating mechanism of ultrasound relaxation at a given temperature. For this purpose, one has only to measure the absorption for the quasi-transverse mode $t_2$ with the polarization vector in the cube face plane in the [100] and [101] directions and find the coefficient $\alpha^*_{Tt2}(\pi/4,0)$. If the inequality $1 \leq \alpha^*_{Tt2}(\pi/4,0) \leq 1.25$ is fulfilled, the isotopic scattering makes the major contribution to the ultrasound relaxation in Si crystals. If $\alpha^*_{Tt2}(\theta_1,0)$ is less than unity and tends to 0.28, the dominant contribution to the ultrasound relaxation in Ge crystals is due to the anharmonic scattering processes.

## 4. Discussion

The analysis demonstrated that the anisotropy of the absorption of quasi-transverse modes in the crystals of the first type is a maximum in diamond. This maximum anisotropy is due mainly to the anisotropy of the third-order elasticity moduli since the anisotropy of the second-order elasticity moduli decreases in going from Ge to Si and then to diamond [17]. The anomalously large anisotropy of the absorption of quasi-transverse modes in the KCl and NaCl crystals is caused by the anisotropy of the anharmonic energy of the crystals. The point is that the absorption of the ultrasound propagating along the cube edges is contributed only by the terms proportional to the third-order elasticity moduli $A_{cub}$ and $\tilde{c}_{155}$ (see formula (17)), which have opposite signs in these crystals (see Table 1). Therefore, their contributions are mutually compensated to a large extent and, as a result, the ultrasound absorption in these directions should be anomalously small as compared to the absorption in Ge where such compensation is not realized. However, the second-order elasticity moduli and, correspondingly, the sound velocity are much smaller in these crystals than those in Ge. For this reason, the coefficient $B_0$ in KCl and NaCl is nearly two orders of magnitude larger than $B_0$ in Ge. These two opposing factors have an effect that the ultrasound absorption in this direction in the KCl, NaCl and Ge crystals is of the same order of magnitude (see Table 2). On the other hand, the absorption of the quasi-transverse modes propagating in directions like [101] and [111] is contributed not only by the terms proportional to the third-order elasticity moduli $A_{cub}$ and $\tilde{c}_{155}$, but also by those proportional to the moduli $\tilde{c}_{111}$ and $\tilde{c}_{112}$ (see formula (10)). Therefore, in crystallographic directions like [101] and [111] such compensation is not realized in the matrix element of the three-phonon scattering processes (see Table 1) and, because of the coefficient $B_0$, the ultrasound absorption in these directions in KCl and NaCl is two orders of magnitude higher than in Ge.

The angular dependences of the ultrasound absorption in GaAs are qualitatively different from the corresponding dependences in the KCl and NaCl crystals and in the majority of cases they are shaped as the dependences in the crystals of the first group. In accordance with the classification by the second-order elasticity moduli, GaAs crystals are referred to the second type: the spectrum of vibration modes in this crystal is qualitatively similar to the spectrum in the KCl and NaCl crystals. The anomalous behavior of the angular dependences of the ultrasound absorption in GaAs is due to the anharmonic energy anisotropy. The third-order elasticity moduli $A_{cub}$ and $\tilde{c}_{155}$ have the same (negative) sign in GaAs crystals, unlike in purely ionic crystals of KCl and NaCl. The elasticity modulus $\tilde{c}_{111}$ has the opposite sign and is nearly 20 times larger in the absolute value than the moduli $A_{cub}$ and $\tilde{c}_{155}$, whereas in KCl these moduli are of the same order of magnitude. As a result, the angle-averaged square of the matrix element of the three-phonon scattering processes decreases approximately four times in moving from directions like [001] to directions like [101]. For this reason, the angular dependences of the ultrasound absorption for quasi-transverse modes $t_2$ and the pure mode $t_1$ with the wave vector in the diagonal plane and the polarization vector perpendicular to this plane are of the same shape in GaAs as in the crystals of the first group. Calculations of the absorption, which were made with the elasticity modulus $\tilde{c}_{111}$ being one order of magnitude smaller, led to dependences qualitatively similar to those in the KCl and NaCl crystals. On the other hand, in the case of the isotropic mode $t_1$ with the wave vector in the cube face plane and the polarization vector perpendicular to this cube face, the absorption is contributed only by the terms proportional to the third-order elasticity moduli $A_{cub}$ and $\tilde{c}_{155}$ (see formula (17)). Consequently, the absorption $\alpha^*_{TLL\,t1}(\theta_1,0)$ in GaAs is inverse to the absorption in the crystals of the first group. Thus, a conclusion follows from the above analysis that the characteristic types of angular dependences of the ultrasound absorption in cubic crystals are determined not only by values of the second-order elasticity moduli, but also by the relation of values and signs of the third-order elasticity moduli, which determine the anharmonic energy of the crystals.

In this study we only consider the volume absorption of transverse long-wave ultrasound. The boundary scattering of phonons is always present in samples of finite dimensions. This scattering is important for relaxation of the acoustic wave momentum at low temperatures: it makes a temperature-independent contribution to the ultrasound absorption. If this contribution is larger than the contribution from the scattering by defects at low temperatures, the specific features of the ultrasound absorption, which are related to the relaxation mechanisms under study, will not be observed in these crystals. The role of the boundary scattering is discussed more comprehensively in [13].

**5. Conclusion**

We have analyzed angular dependences of the transverse ultrasound absorption in cubic crystals with positive and negative anisotropies of the second-order elasticity moduli. The effect of the cubic anisotropy on the spectrum and polarization vectors of vibration modes was taken into account. Two most important cases – wave vectors of phonons are in the cube face plane or the diagonal planes – were considered in terms of the anisotropic continuum model. Known values of the second- and third-order elasticity moduli were used to calculate parameters determining the ultrasound absorption in the crystals under study. The anisotropy of the transverse ultrasound absorption in the presence of the competition between the defect and anharmonic scattering processes was analyzed. It was shown that the angular dependences of the transverse ultrasound absorption are inverse for the anharmonic scattering processes and the scattering by defects. Therefore, the dominant mechanism of ultrasound relaxation can be determined by analyzing the ultrasound absorption anisotropy in cubic crystals. The analysis demonstrated that the angular dependences of the transverse ultrasound absorption in cubic crystals with positive (Ge, Si and diamond) and negative (KCl, NaCl) anisotropies of the second-order elasticity moduli differ qualitatively for the Landau-Rumer relaxation mechanism. The absorption anisotropy in the crystals of the second type is one order (NaCl) and two orders of magnitude (KCl) larger than it is in Ge, Si and diamond. It was shown that the increase in the absorption by four orders of magnitude in GaAs as compared to the absorption in Ge or InSb is connected with anomalously small values of the second-order elasticity moduli, which are one order of magnitude smaller than in Ge and InSb crystals. The contributions of longitudinal components of quasi-transverse vibrations to the absorption of quasi-transverse modes were estimated. They are in qualitative agreement with results of the analysis of polarization vectors performed in [17].

**Acknowledgment**

The authors wish to thank A.P. Tankeyev, A.V. Inyushkin, G.D. Mansfeld, V.A. Volkov and S.G. Alekseyev for discussion of the results and critical remarks.

This study was supported under the Presidium RAS program No. 12, the program for support of leading scientific schools (grant No. НШ 5869.2006.2), and the Russian Science Support Foundation.